# A Simulation Study of Bandit Algorithms to Address External Validity of Software Fault Prediction

Teruki Hayakawa
*Kindai University*
Higashiosaka, Japan
1610370891g@kindai.ac.jp

Masateru Tsunoda
*Kindai University*
Higashiosaka, Japan
tsunoda@info.kindai.ac.jp

Koji Toda
*Fukuoka Institute of Technology*
Fukuoka, Japan
toda@fit.ac.jp

Keitaro Nakasai
Nara Institute of Science and Technology
Ikoma, Japan
nakasai.keitaro.nc8@is.naist.jp

Kenichi Matsumoto
*Nara Institute of Science and Technology*
Ikoma, Japan
matumoto@is.naist. jp

***Abstract*—Various software fault prediction models and techniques for building algorithms have been proposed. Many studies have compared and evaluated them to identify the most effective ones. However, in most cases, such models and techniques do not have the best performance on every dataset. This is because there is diversity of software development datasets, and therefore, there is a risk that the selected model or technique shows bad performance on a certain dataset. To avoid selecting a low accuracy model, we apply bandit algorithms to predict faults. Consider a case where player has 100 coins to bet on several slot machines. Ordinary usage of software fault prediction is analogous to the player betting all 100 coins in one slot machine. In contrast, bandit algorithms bet one coin on each machine (i.e., use prediction models) step-by-step to seek the best machine. In the experiment, we developed an artificial dataset that includes 100 modules, 15 of which include faults. Then, we developed various artificial fault prediction models and selected them dynamically using bandit algorithms. The Thomson sampling algorithm showed the best or second-best prediction performance compared with using only one prediction model.**

## I. INTRODUCTION

In project management, planning strategies are based on the prediction of the project. For example, the software testing plan is made based on software fault prediction. In previous studies, various software fault prediction models and techniques, such as feature selection [4], have been proposed for building software. Many studies compared and evaluated various models [2] and techniques to identify the most efficient ones. Such evaluations considered various datasets and showed which model and technique is the most effective on average.

However, in most cases, these models do not have the highest performance on every dataset (e.g., [2][4]). This is because there is diversity of software development datasets, and therefore, there is a risk that the selected model or technique shows bad performance on a certain dataset. D'Ambros et al. [2] pointed out that external validity in fault prediction is still an open problem. This is because practitioners use only one fault prediction model during software testing. In addition, if two new fault prediction models have been developed, and they have not been compared, it is not clear which model should be used.

To avoid selecting a low accuracy model, we apply bandit algorithms. Although bandit algorithms [9] are relatively classical, they have been recently utilized in various fields, such as website optimization [10]. Bandit algorithms are often explained through an analogy with slot machines. Assume that a player has 100 coins to bet on several slot machines, and he/she wants to maximize the reward. If the player does not know the bandit algorithm, he/she might select only one slot machine and bet all 100 coins on that machine. In contrast, bandit algorithms require the player to bet one coin on each slot machine to seek the best one.

Fig.1 shows the difference of test phase between existing and our approach. On existing approach, fault prediction models are built and evaluated on test planning phase. In test execution phase, the built model is used. The details are shown in Table I. In the table, candidates of prediction model (i.e., SVM and hard voting (ensemble method)) are evaluated based on evaluation criteria such as AUC. However, the selected model (i.e., hard voting) is not work well on test execution phase (The prediction is not accurate). In the table, gray cells indicates wrong predictions.

In contrast, on our approach, only the model building is performed on test planning phase, and model evaluation is done in test execution phase. That is, both evaluating model and using model are performed in test execution phase repeatedly. Table II shows the detail of the phase. Intuitively speaking, all prediction models are evaluated when test of each module is finished, and based on the result (i.e., whether the prediction is correct or not), the reward is set to each model. Based on the average reward (i.e., accuracy), used model is dynamically

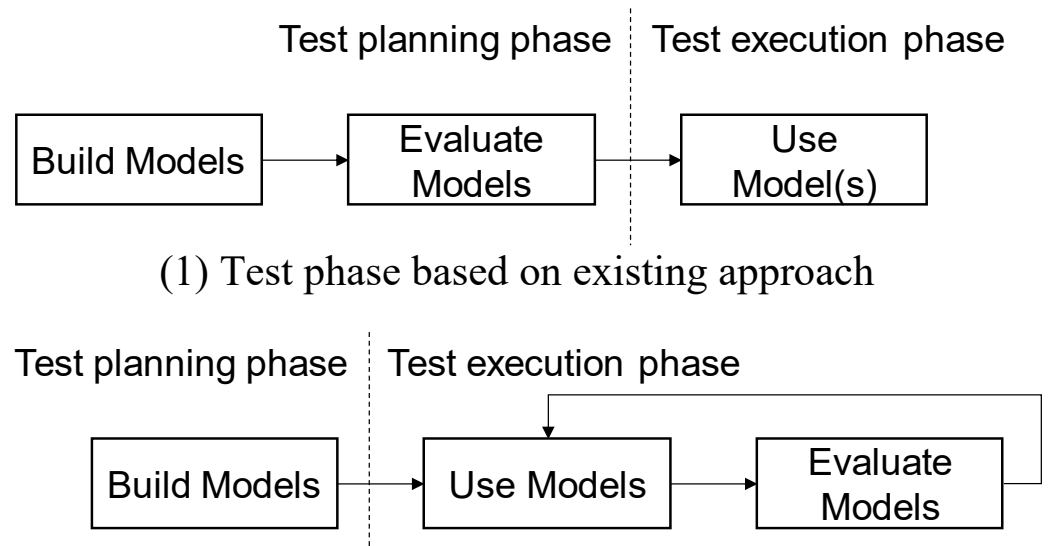

(2) Test phase based on Bandit Algorithms

Fig. 1. Difference of test phase between existing and our approach

TABLE I. DETAIL OF TEST PHASE BASED ON EXISTING APPROACH

(1) Model evaluation based on past data on test planning phase

| Module | Prediction by hard voting | Prediction by SVM | | Found faults? |
|---|---|---|---|---|
| Learn1.java | FP (Fault-prone) | NFP (not-fault-prone) | | Yes |
| Learn2.java | FP | NFP | | Yes |
| Learn3.java | FP | NFP | | Yes |
| Learn4.java | NFP | FP | | No |
| Learn5.java | NFP | FP | | No |
| Learn6.java | NFP | FP | | No |
| … | … | … | | … |
| **AUC** | 0.9 | 0.5 | | |

(2) Using the model on test execution phase

| Module | Prediction by hard voting | Found faults? |
|---|---|---|
| Test1.java | FP | Yes |
| Test5.java | FP | No |
| Test2.java | NFP | Yes |
| Test3.java | NFP | Yes |
| Test4.java | NFP | No |
| Test6.java | FP | No |
| … | … | … |
| **AUC** | 0.5 | |

selected. In the table, SVM was dynamically selected after software testing of Test5.java is finished. Details of the procedure is explained in section III using Table IV.

## II. BANDIT ALGORITHMS

As explained in section 1, Bandit algorithms are often explained through an analogy with slot machines, which are called arms in bandit algorithms. The following is the simplest bandit algorithm:

- The player bets one coin on one arm.
- **Exploration phase**: When the average reward of the currently selected arm is lower than that of others, the player selects another arm with higher average reward. This action is aimed at finding the arm with the highest reward, and therefore, it is referred to as exploration.
- **Exploitation phase**: When the average reward of the current arm is the highest (or equal), the player keeps playing on it. This action is aimed at exploiting the highest reward arm, and therefore, it is referred to as exploitation.

Table III shows an example of the algorithm. The following is a detailed explanation of the table. Initially, the average reward of each arm is zero.

1) Arm A is selected randomly. When the reward of arm A is -1, its average reward becomes -1.0.
2) Arm B is selected because its average reward is higher than that of A. When the reward of arm B is 1, its average reward becomes 1.0.
3) Arm B is selected again because its average reward is higher than that of A.

The first and third trial are regarded as the exploitation phase, and the second trial is regarded as the exploration phase.

**Epsilon greedy strategy**: The above algorithm does not always select the best arm, as shown in Table IV. The following is a detailed explanation of the table.

1) Arm A is selected randomly. When the reward of arm A is 1, its average reward becomes 1.0.
2) Arm A is selected again, because its average reward is higher than that of B. When the reward of arm A is -1, its average reward becomes 0.
3) Arm A is selected again because its average reward is equal to that of B.
4) Arm B is selected because its average reward is higher than that of A. When the reward of arm B is -1, its average reward becomes -1.0.
5) After the fourth trial, arm A is always selected, because its average reward is always higher than that of B.

In the above situation, arm A is selected repeatedly, even if the actual average reward of arm B is larger. The arm selection is dependent on the first reward (i.e., the first reward of machine A is positive, and that of machine B is negative).

The epsilon greedy strategy can avoid such a situation. The strategy selects arms through the following algorithm.

- The best arm is selected with probability $1 - \varepsilon$ ($0 \leq \varepsilon \leq 1$) for the exploitation phase based on the average reward of each arm.
- One of the arms is selected with probability $\varepsilon$ for the exploration phase.

When the value of $\varepsilon$ is 0, arms are always selected based on the average reward of each arm. In contrast, when the value of $\varepsilon$ is 1, arms are always selected randomly. Table V shows an example of the epsilon greedy strategy when the value of $\varepsilon$ is larger than zero. Table V shows only the fifth and sixth trials because the others are the same as in Table IV:

5) Arm B is randomly selected. When the reward of arm B is 1, its average reward becomes 0.
6) Arm B is selected because its average reward is higher than that of A.

TABLE II. USING AND EVALUATIONG MODELS ON TEST EXECUTION PHASE BASED ON BANDIT ALGORITHMS

| Module | Prediction by hard voting | Is Hard Voting correct? =Reward (Yes: 1,No: 0) | Prediction by SVM | Is SVM correct? =Reward (Yes: 1,No: 0) | Found faults? | Prediction and used model based on average reward |
|---|---|---|---|---|---|---|
| Test1.java | FP | 1 | FP | 1 | Yes | NFP (Hard Voting) |
| Test5.java | FP | 0 | NFP | 1 | No | FP (Hard Voting) |
| Test2.java | NFP | 0 | FP | 1 | Yes | FP (SVM) |
| Test3.java | NFP | 0 | FP | 1 | Yes | FP (SVM) |
| Test4.java | NFP | 0 | NFP | 1 | No | NFP (SVM) |
| Test6.java | FP | 0 | NFP | 1 | No | NFP (SVM) |
| … | … | … | … | … | … | … |

TABLE III. EXAMPLE OF BANDIT ALGORITHM

| # of trial | Selected arm | Reward | Avg. reward of A | Avg. reward of B |
|---|---|---|---|---|
| 1 | A | -1 | -1.0 | 0 |
| 2 | B | 1 | -1.0 | 1.0 |
| 3 | B | 1 | -1.0 | 1.0 |

TABLE IV. EXAMPLE TO FAIL ARM SELECTION

| # of trial | Selected arm | Reward | Avg. reward of A | Avg. reward of B |
|---|---|---|---|---|
| 1 | A | 1 | 1.0 | 0 |
| 2 | A | -1 | 0 | 0 |
| 3 | A | -1 | -0.33 | 0 |
| 4 | B | -1 | -0.33 | -1.0 |
| 5 | A | -1 | -0.5 | -1.0 |
| 6 | A | -1 | -0.6 | -1.0 |

TABLE V. EXAMPLE OF EPSILON GREEDY STRATEGY

| # of trial | Selected arm | Reward | Avg. reward of A | Avg. reward of B |
|---|---|---|---|---|
| 5 | B | 1 | -0.33 | 0 |
| 6 | B | 1 | -0.33 | 0.33 |

After the sixth trial, the cumulated reward in Table IV is -4, and that in Table V is 0, which shows the effectiveness of the epsilon greedy strategy is in this situation.

**Related work**: In other fields, bandit algorithms are used to optimize machine learning models. Li et al. [5] applied bandit algorithms to the models to adjust hyperparameters. However, the method proposed in the study assumes to apply bandit algorithms to learning data, and therefore, it cannot be applied during software testing.

A/B testing [10] can be applied during software testing, because it has the exploitation and exploration phases. A/B testing performs exploration phase to settle the best arm before exploitation phase. The exploitation phase is longer than the exploration phase, and they are clearly distinguished. On the exploration phase, each arm are selected at same probability. Therefore, A/B testing is not considered to be better choice, compared with bandit algorithms [10].

## III. PREDICTION EXPLORATION DURING SOFTWARE TESTING

The common usage of software fault prediction algorithms explained in section 1, is analogous to betting all 100 coins in one slot machine. The comparisons and evaluations performed in previous studies can be regarded as the exploration phase, and the testing of software using the selected models only includes the exploitation phase.

In this study, we apply the bandit algorithm to introduce the exploration phase on software testing. We use multiple fault prediction models, and regard the prediction results as arms.

Table VI explains the fault prediction process of the bandit algorithm. Note that Table IV explains detail process of Table II. In the table, gray cells indicate tested modules that are selected by the prediction. The following are the premises of the example.

- Modules Test1.java–Test3.java include faults, and Test4.java–Test6.java do not.

TABLE VI. APPLYING BANDIT ALGORITHMS TO PREDICT DEFECTS

(1) First step

| Model A (e.g., hard voting) | | | Model B (e.g., SVM) | | |
|---|---|---|---|---|---|
| Module | Pred. | Reward | Module | Pred. | Reward |
| Test1.java | FP | 1 | Test1.java | FP | 1 |
| Test5.java | FP | | Test2.java | FP | |
| Test6.java | FP | | Test3.java | FP | |
| Test2.java | NFP | | Test4.java | NFP | |
| Test3.java | NFP | | Test5.java | NFP | |
| Test4.java | NFP | | Test6.java | NFP | |
| Avg. reward | | 1 | Avg. reward | | 1 |

(2) Second step

| Model A (e.g., hard voting) | | | Model B (e.g., SVM) | | |
|---|---|---|---|---|---|
| Module | Pred. | Reward | Module | Pred. | Reward |
| Test1.java | FP | 1 | Test1.java | FP | 1 |
| Test5.java | FP | -1 | Test2.java | FP | |
| Test6.java | FP | | Test3.java | FP | |
| Test2.java | NFP | | Test4.java | NFP | |
| Test3.java | NFP | | Test5.java | NFP | 1 |
| Test4.java | NFP | | Test6.java | NFP | |
| Avg. reward | | 0 | Avg. reward | | 1 |

(3) Third step

| Model A (e.g., hard voting) | | | Model B (e.g., SVM) | | |
|---|---|---|---|---|---|
| Module | Pred. | Reward | Module | Pred. | Reward |
| Test1.java | FP | 1 | Test1.java | FP | 1 |
| Test5.java | FP | -1 | Test2.java | FP | |
| Test6.java | FP | | Test3.java | FP | |
| Test2.java | NFP | | Test4.java | NFP | |
| Test3.java | NFP | | Test5.java | NFP | 1 |
| Test4.java | NFP | | Test6.java | NFP | |
| Avg. reward | | 0 | Avg. reward | | 1 |

- The column "Pred." is the predicted result, and when its value is "FP (fault prone)", the model predicted that the module includes faults. When its value is "NFP (not-fault prone)", it predicted that the module does not include faults.
- Before testing, for each prediction model, prediction results should be sorted in descending order based on the results, to test faulty modules first. When the modules are predicted as non-faulty, the number of test cases for them is set as smaller [6]. That causes overlook of faults, and test results (i.e., actual values) becomes unreliable.
- When the test and predicted results are the same, the reward is 1, and when they are different, the reward is -1. The average reward of each model is recalculated after every test.

The following is a detailed explanation of the table.

1) Model A is selected, and module Test1.java is tested based on the prediction. In this case, the test and prediction results are the same, and the average reward of model A becomes 1. Simultaneously, we evaluate the result of model B, and its average reward turns out to be 1. This is because we know the results of all prediction models before testing.
2) Model A is selected, and module Test5.java is tested. The test and prediction results are different, and the average reward of model A becomes 0. In contrast, the result of model B is correct, and therefore, the average reward of model B remains 1.
3) Model B is selected because its average reward is higher than that of A. Although module Test1.java has the highest

TABLE VII. AN EXAMPLE OF ARTIFICAL DATASET USED IN EXPERIMENT

| Module | Prediction by Model1 | … | Prediction by Model4 | Found faults? |
|---|---|---|---|---|
| Test1.java | NFP | | NFP | No |
| Test2.java | FP | | NFP | No |
| Test3.java | FP | | NFP | No |
| Test4.java | NFP | | FP | Yes |
| Test5.java | NFP | … | FP | Yes |
| Test6.java | NFP | | FP | Yes |
| … | … | … | … | … |
| **AUC** | 0.59 | … | 0.80 | |

prediction result of model B, it not tested because it was already tested. Therefore, module Test2.java is tested based on the prediction of B.

**Difference with normal bandit algorithms**: As explained above, we can evaluate all arms (i.e., prediction models) at each trial during software testing because we know the prediction results of all models before testing. The calculation is very different from that of normal bandit algorithms, and has the following advantages:

- Convergence of average reward of each arm is faster than that of normal bandit algorithms.
- The convergence speed is not significantly affected by the number of arms. This is because we can evaluate all arms simultaneously.
- The selection of a random arm is not strictly necessary. This is because the average reward of all arms can be calculated, even if some arms are not selected.

**Applicability of bandit algorithms to prediction models**: The following are the reasons why we can apply bandit algorithms to fault prediction models.

- Predictions are repeatedly performed step-by-step. If software testing is performed in a short time or only once during a software development project, bandit algorithms cannot be applied.
- Prediction models are not changed during software testing. This is because the characteristics of the software (and software) do not change significantly during software testing (i.e., if the model and its accuracy change during software testing, the average reward result is unreliable).
- We can easily check the correctness of the fault predictions during software testing by comparing the test results with the prediction results. This is because the test results are often input to issue tracking system (i.e., if test results are not recorded, we cannot compare the results and, as a result, we cannot set the rewards during testing).

If the predictions satisfy the above conditions, we may apply bandit algorithms to other prediction models.

**Difference with ensemble techniques**: While both bandit algorithms and ensemble techniques [7] use multiple prediction models, ensemble techniques do not include the exploration phase during software testing. The prediction results of ensemble techniques are regarded as one of the arms, and therefore, bandit algorithms and ensemble techniques can be used together. For example, we can dynamically select hard and soft voting during test execution phase, using bandit algorithms.

**Needed effort for bandit algorithms**: To apply bandit algorithms in fault prediction models during software testing, the following tasks are required.

1) Making multiple prediction models.
2) Recording testing results.
3) Comparing prediction and testing results.
4) Selecting a model based on the average reward.

Task 1 is performed only once, and task 2 is generally performed during testing even when we do not apply bandit algorithms. Tasks 3 and 4 are performed automatically using software. While additional tasks are required, when applying bandit algorithms the risk of using low accuracy prediction models can be suppressed. Unlike bandit algorithms, conventional models overlook fault modules, which increased the cost of software testing.

## IV. EXPERIMENT

**Overview**: In the experiment, we developed an artificial dataset and predictions to evaluate the effectiveness of bandit algorithms. The experiment had the following components.

- **Dataset**: We developed an artificial dataset that includes 100 modules, 15 of which contain faults. This is because the datasets used for fault prediction are often imbalanced. To control the characteristics of dataset, artificial dataset is appropriate [8].
- **Models (arms)**: We set models (arms) with various accuracies, high accuracy, and low accuracy. We prepared various artificial dataset as shown in Table VII. In the table, the prediction is not based on actual prediction model, but randomly generated. This is because the results of bandit algorithms depend on the prediction results only. The accuracy of each model is shown in Table VIII, IX, and X, respectively.
- **Reward**: We set the reward to 1 when the prediction and test results were the same, and to -1 when they were different.

TABLE VIII. VARIOUS ACCURACY MODEL SET

| Model 1 | Model 2 | Model 3 | Model 4 |
|---|---|---|---|
| 0.59 | 0.70 | 0.77 | **0.80** |

TABLE IX. HIGH ACCURACY MODEL SET

| Model 1 | Model 2 | Model 3 | Model 4 |
|---|---|---|---|
| 0.70 | 0.78 | 0.82 | **0.88** |

TABLE X. LOW ACCURACY MODEL SET

| Model 1 | Model 2 | Model 3 | Model 4 |
|---|---|---|---|
| 0.50 | 0.53 | 0.54 | **0.59** |

TABLE XI. ACCURACY OF BANDIT ALGORITHMS USING THE MODELS SHOWN IN TABLE V

| ε = 0 | ε = 0.1 | UCB | TS |
|---|---|---|---|
| 0.81 | 0.81 | 0.81 | **0.81** |

TABLE XII. ACCURACY OF BANDIT ALGORITHMS USING THE MODELS SHOWN IN TABLE VI

| ε = 0 | ε = 0.1 | UCB | TS |
|---|---|---|---|
| 0.86 | 0.86 | 0.86 | **0.91** |

TABLE XIII. ACCURACY OF BANDIT ALGORITHMS USING THE MODELS SHOWN IN TABLE VII

| ε = 0 | ε = 0.1 | UCB | TS |
|---|---|---|---|
| 0.52 | 0.52 | 0.55 | **0.56** |

- **Bandit algorithms**: We applied the epsilon greedy strategy, and set ε to 0 and 0.1. Additionally, we used the upper confidence bound (UCB) [10] and Thompson sampling [3]. UCB selects arms whose upper confidence bounds are the largest at 1 / *n* significance level (*n* is the number of playtimes). Thompson sampling (TS) selects arms whose rewards are the highest based on posterior probability.
- **Evaluation criteria**: We used the area under the curve (AUC) to evaluate the fault prediction model (e.g., [2][4]). AUCs are derived by averaging each iteration result, as explained below.
- **Procedure**: We repeated the experiment ten times. This is because bandit algorithms select arms (models) randomly, and therefore, the results of the evaluation are different for each iteration.

**Results-algorithm evaluation**: The prediction accuracy of bandit algorithms are shown in Tables XI–XIII. In each table, bold numbers indicate the highest accuracy. The results are as follows.

- Among the algorithms, Thompson sampling showed the highest accuracy (Tables XI–XIII). Note that it had only slightly higher accuracy in Table XI.
- The Thompson sampling algorithm showed the highest accuracy when using the various- and high-accuracy model sets (Tables VIII and IX), and it showed the second-highest accuracy when using the low-accuracy model set (Table X).

The results suggest that, when Thompson sampling is used, we can avoid using the lowest accuracy prediction. Moreover, we can use the second-highest accuracy prediction at least.

**Results-comparison with existing approach (baseline)**: We compared the result with existing approach (i.e., baseline). Based on existing approach, models are not evaluated when test execution phase as shown in Fig. 1. That is, we may select the lowest accuracy model or the highest accuracy model (e.g., in Table VII, the former is model 1, and the latter is model 4). Even if we set the baseline as the highest accuracy model in Tables VIII-X, our approach with Thompson sampling is better than the baseline in two out of three cases, as shown in Tables XI and XII. If we set the baseline as the lowest accuracy model in Tables VIII-X, our approach is better than all cases.

**Threats to validity**: The effect of bandit algorithms depends on the order of prediction (i.e., reward evaluation). For example, if the prediction results of model B in Table VI were incorrect after the third step, the model would be used in some steps, possibly leading to lower prediction accuracy. In the experiment, we made various predictions artificially, and therefore, the probability of selection was not large.

We did not use actual fault prediction models, but used artificial predictions. Bowes et al. [1] showed that different fault prediction models find different faults (i.e., not correlated). Therefore, artificial predictions is considered to be practical.

## V. Conclusion

We applied bandit algorithms to software fault predictions to avoid selecting low-accuracy prediction models. Bandit algorithms are applied in various fields, such as website optimization. To the best of our knowledge, this is the first instance of a bandit algorithm being applied to software fault prediction.

In the experiment, we prepared an artificial dataset and three prediction model sets. Each model set included four artificial prediction models. We applied four bandit algorithms (epsilon greedy strategy (ε = 0 and 0.1), UCB, and Thompson sampling) and evaluated the prediction accuracy using AUC. Thompson sampling showed the highest or the second-highest prediction performance, compared with prediction models included in each model set.

The results confirmed that bandit algorithms can prevent the selection of low accuracy fault prediction models and achieve better prediction performance than conventional algorithms. In the future, we will apply bandit algorithms to actual datasets and prediction models.

## Acknowledgments

This research was partially supported by the Japan Society for the Promotion of Science (JSPS) [Grants-in-Aid for Scientific Research (A) (No.17H00731)].